\documentclass[notitlepage,prl,byrevtex,superscriptaddress,12pt]{revtex4}
\usepackage{amsfonts}
\usepackage{amsmath}
\usepackage{amssymb}
\usepackage{graphicx}

\begin{document}

\title{A novel and precise time domain description of MOSFET low frequency
noise due to random telegraph signals}
\author{Roberto da Silva}
\affiliation{Instituto de Informatica\\
Universidade Federal do Rio Grande do Sul\\
Av Bento Gon\c{c}alves, 9500\\
91501-970 - Porto Alegre RS, Brasil}
\author{Gilson Inacio Wirth}
\affiliation{Escola de Engenharia\\
Universidade Federal do Rio Grande do Sul\\
Osvaldo Aranha, 103, 2%
%TCIMACRO{\U{ba} }%
%BeginExpansion
${{}^o}$
%EndExpansion
andar\\
90035-190, Porto Alegre, RS, Brasil }
\author{Lucas Brusamarello}
\affiliation{Instituto de Informatica\\
Universidade Federal do Rio Grande do Sul\\
Av Bento Gon\c{c}alves, 9500\\
91501-970 - Porto Alegre RS, Brasil}

\begin{abstract}
Nowadays, random telegraph signals play an important role in integrated
circuit performance variability, leading for instance to failures in memory
circuits. This problem is related to the successive captures and emissions
of electrons at the many traps stochastically distributed at the
silicon-oxide (S$_{\text{i}}$-S$_{\text{i}}$O$_{\text{2}}$) interface of MOS
transistors. In this paper we propose a novel analytical and numerical
approach to statistically describe the fluctuations of current due to random
telegraph signal in time domain. Our results include two distinct
situations: when the density of interface trap density is uniform in energy,
and when it is an u-shape curve as prescribed in literature, here described
as simple quadratic function. We establish formulas for relative error as
function of the parameters related to capture and emission probabilities.
For a complete analysis experimental u-shape curves are used and compared
with the theoretical aproach.
\end{abstract}

\maketitle

\begin{figure}[th]
\centering \includegraphics[width=0.49\columnwidth]{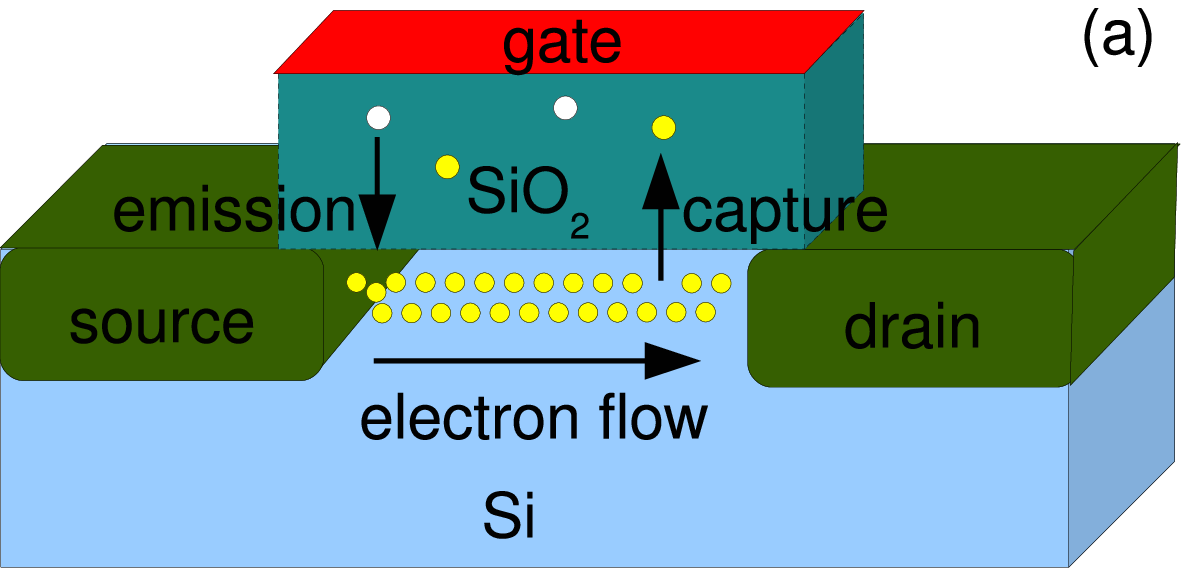} %
\includegraphics[width=0.49\columnwidth]{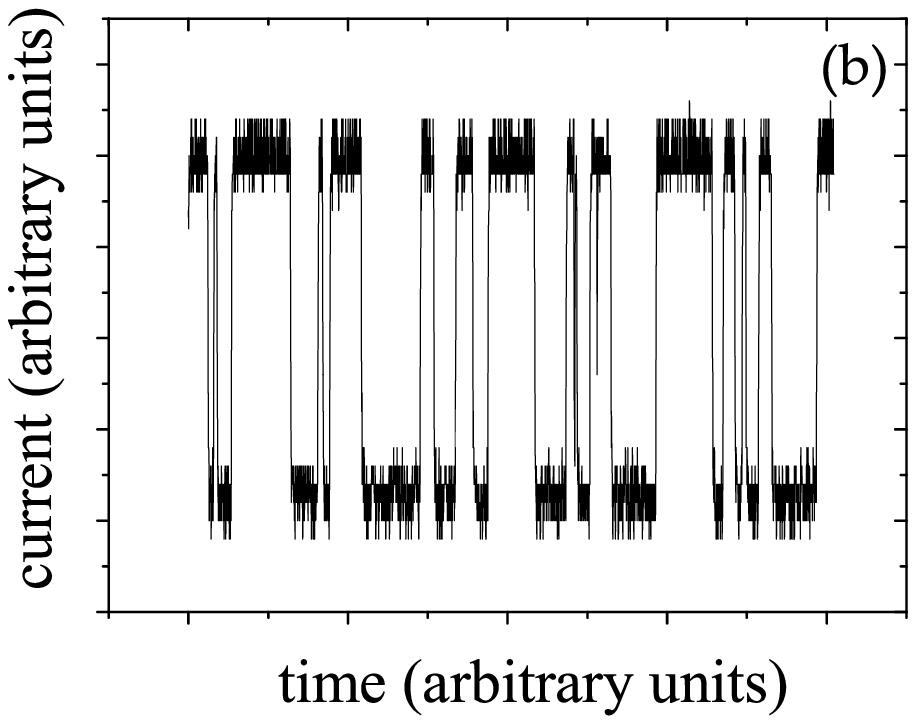}
\caption{\textbf{Plot a}: Physical mechanism for the generation of LF-noise;
\textbf{Plot b}: This figure shows the effects caused by only one trap
according to physical mechanism described in plot a. A discontinuity in the
electrical current corresponds to the absorption or emission of an electron.
}
\label{mechanism_LF-noise}
\end{figure}

Low frequency (LF) noise is a performance limiting factor for deep
sub-micron CMOS devices \cite{KU1989}. In these devices, LF noise is
dominated by multiple Random Telegraph Signals (RTS) (see \cite{M1954} for a
discussion on RTS in S$_{\text{i}}$-S$_{\text{i}}$O$_{\text{2}}$
interfaces). The MOS transistor employed in analog and digital designs
drives a current $I_{DS}$ which is a function of the potential differences $%
V_{DS}$ and $V_{GS}$ applied respectively between drain and source and gate
and source\cite{Shew87}. RTS noise is due to succeeding electron capture and
emission by a number of $N_{tr}\ $traps distributed according to a Poisson
distribution at the S$_{i}$-S$_{i}$O$_{2}$ interface, as it can be observed
in plot \ref{mechanism_LF-noise} (a). This phenomena causes oscillations in
the transistor current $I_{DS}$ over time, even with $V_{DS}$ and $V_{GS}$
constant. A typical variation in the current caused by one single trap is
shown in figure \ref{mechanism_LF-noise} (b).

Noise performance may strongly vary between different devices on a same
chip, and even between different operation points of a single transistor.
Recent results \cite{CGSLBV2008} claim these effects responsible for fails
in flash memories. For future technologies, RTS may become relevant also for
the performance of combinational digital gates, as inverters and other logic
gates. The propagation delay of a gate depends on the capability of its
transistors to drive current. A smaller current driven by the transistor
means larger propagation delay, which may lead to timing violations
(failures). Hence, the variation in transistor drive current due to RTS may
lead to circuit failures in future technology generations, and statistical
modeling of random telegraph signal is demanded. But until now statistical
models for RTS focused on frequency domain, for instance refer to \cite%
{dasilva_2008_appropriate} for a statistical RTS model in frequency domain.
This is suitable for analog circuits, whose design and analysis is performed
in frequency domain. However, for digital circuits an appropriate time
domain statistical analysis is needed, since these circuits are analyzed and
designed using time domain metrics. So, as RTS is becoming a major issue on
upcoming digital integrated circuits, an approppriate RTS noise model for
time domain is mandatory for the correct modeling of the behavior of fugure
digital circuits.

Aiming to address this issue, this work presents for the first time in
literature a comprehensive model for RTS in time domain. We derive
analytical expressions for the average and relative error of the transistor
current. We also analyze the joint contribution of all the traps to the
transistor current, instead of analyzing only the current fluctuation due to
one single trap, which is more suitable for analyzing the total current
fluctuation of the transistor. Our model considers the density of states in
the band-gap analyzing three possible situations:

\begin{enumerate}
\item The density of states is uniform in the band-gap range. This is based
on previous results from literature (see for example \cite{KU1989})

\item The density of states curve measured from experimental data, which was
published in \cite{WY1990} and shows that the density of states follows an
u-shape curve. We performed a polynomial fit with a 8-degree polynomial
which fits very well with the experimental data.

\item The density of states modeled as a quadratic function in the band-gap
range. This is an approximation for the experimental data of \cite{WY1990},
which is an u-shape curve.
\end{enumerate}

In the random telegraph signals, $\sigma _{i}=0,1$ denotes the state of the
i-th trap ($0=$empty or $1=$occupied), the Fermi-Dirac statistics governs
the probability of transition:
\begin{eqnarray*}
\Pr (\sigma _{i} &=&0\rightarrow \sigma _{i}=1)dt=\frac{dt}{10^{p_{i}}\left[
1+\exp (-q_{i})\right] }=\frac{dt}{\tau _{c}^{(i)}} \\
\Pr (\sigma _{i} &=&1\rightarrow \sigma _{i}=0)dt=\frac{dt}{10^{p_{i}}\left[
1+\exp (q_{i})\right] }=\frac{dt}{\tau _{e}^{(i)}}
\end{eqnarray*}%
where $\tau _{c}^{(i)}$ and $\tau _{e}^{(i)}$ are respectively the average
time of capture and emission and $q_{i}=(E_{T}^{(i)}-\mu )/k_{B}T$ where $%
E_{T}^{(i)}\ $is the energy ( within the band-gap) of the $i-$th trap, $\mu $
is the fermi-level energy, $k_{B}=1.3806568\times 10^{-23}Jk_{B}^{-1}$ the
Boltzmann constant and $T$ is temperature. In a first model $-Q<q_{i}<Q$ is
considered a uniform random variable and $\tau _{c}^{(i)}$, $\tau _{e}^{(i)}$
are identically distributed, i.e., $\left\langle \tau
_{c}^{(i)}\right\rangle =\left\langle \tau _{c}\right\rangle $ and $%
\left\langle \tau _{e}^{(i)}\right\rangle =\left\langle \tau
_{e}\right\rangle \ $for $i=1,2,...,N_{tr}$.

Here, $p_{i}$ is also a random uniform variable within an interval $p_{\min
}<p_{i}<p_{\max }$ and in this case in the frequency domain, we can
establish an important connection. It is not difficult to show (see for
example \cite{M1954,SWB2006,WSKTB2005,WSB2007}) that the power spectrum
density correponding to the noise from the $i$-th trap is a Lorentzian
function $S_{i}(f_{i})=(A_{i}^{2}/f_{i})\left[ 1+(f/f_{i})^{2}\right] ^{-1}$
where $f_{i}$ $=1/\tau _{c}^{(i)}+1/\tau _{e}^{(i)}\ $is the corner
frequency corresponding to the trap and $A_{i}$ is its amplitude. From that
we conclude $f_{i}=10^{-p_{i}}$ and due to this $f_{i}$ is uniformly
distributed in a $\log _{10}$ scale resulting in a probability distribution $%
h(f_{i})=\left[ \ln 10\ (p_{\max }-p_{\min })\ f_{i}\right] ^{-1}$ for the
corner frequencies (this assumption will be used from now on in this paper).

From this approach, we can calculate
\begin{equation*}
\Pr (\sigma _{i}=1)=\frac{1}{\tau _{e}}/(\frac{1}{\tau _{c}}+\frac{1}{\tau
_{e}})=\frac{\tau _{c}}{\tau _{c}+\tau _{e}}
\end{equation*}%
and thus the average current due to a single trap is computed by

\begin{equation*}
\begin{array}{lll}
\overline{I_{i}} & = & \delta_{i}\cdot\Pr(\sigma_{i}=1)+0\cdot\Pr(\sigma
_{i}=0) \\
&  &  \\
& = & \frac{\tau_{c}^{(i)}}{\tau_{c}^{(i)}+\tau_{e}^{(i)}}\delta_{i}%
\end{array}%
\end{equation*}

The amplitudes $\delta _{i}$ are also random variables and our results will
be dependent on its first and second moments, respectively $\left\langle
\delta \right\rangle $ and $\left\langle \delta ^{2}\right\rangle $. The
next step is to consider the contribution of all traps, because by doing so
one can compute the fluctuation of the current passing through the channel
of the MOS transistor. The average current, taking in consideration all
random sources, is written as:$\ $
\begin{equation*}
\begin{array}{lll}
\left\langle \overline{I}\right\rangle =\left\langle \overline{I_{i}}%
\right\rangle & = & \left\langle \sum\limits_{i=1}^{N_{tr}}\overline{I_{i}}%
\right\rangle _{N_{tr},\tau _{c},\tau _{e}} \\
&  &  \\
& = & \frac{\left\langle \delta \right\rangle }{2Q}\left(
\sum\limits_{N_{tr}=0}^{\infty }N_{tr}\dfrac{N^{N_{tr}}e^{-_{N}}}{N_{tr}!}%
\right) \int\limits_{-Q}^{Q}\frac{1+e^{-q}}{2(1+\cosh q)}dq \\
&  &  \\
& = & \frac{N\ \left\langle \delta \right\rangle }{2Q}\ \left[ 2Q+\ln \frac{%
\left( e^{-Q}+1\right) }{\left( e^{Q}+1\right) }\right] \\
&  &  \\
& = & \frac{N\ \left\langle \delta \right\rangle }{2Q}\ \left[ 2Q+\ln e^{-Q}%
\right] =\frac{N\ \left\langle \delta \right\rangle }{2}%
\end{array}%
\end{equation*}%
where $N=\ c\ (p_{\max }-p_{\min })$ and $c=\ln 10\ N_{dec}W\ L$, where $W$
and $L$ are the device dimensions and $\rho =\ln 10\ N_{dec}$ is the density
of traps by area unit and by decade frequency in log-scale. So, we can
conclude that $\left\langle \overline{I}\right\rangle $ is constant, i.e.,
it does not depend on $Q$. This conclusion motivates the investigation for
superior moments of $\overline{I}$.

Similarly we can compute the second moment:%
\begin{equation*}
\begin{array}{lll}
\left\langle I^{2}\right\rangle & = & \left\langle \left(
\sum\limits_{i=1}^{N_{tr}}\overline{I_{i}}\right) ^{2}\right\rangle
_{N_{tr},\tau _{c},\tau _{e}} \\
&  &  \\
& = & \left\langle \sum\limits_{i=1}^{N_{tr}}\overline{I_{i}}%
^{2}+\sum\limits_{i=1}^{N_{tr}}\sum\limits_{\substack{ j=1  \\ i\neq j}}%
^{N_{tr}}\overline{I_{i}}\overline{I_{j}}\right\rangle _{N_{tr},\tau
_{c},\tau _{e}} \\
&  &  \\
& = &
\begin{array}{l}
\sum\limits_{N_{tr}=0}^{\infty }\dfrac{e^{-_{N}}N^{N_{tr}}}{N_{tr}!}\left(
N_{tr}\left\langle \overline{I}^{2}\right\rangle _{\tau _{c},\tau
_{e}}\right. \\
\left. +N_{tr}(N_{tr}-1)\left\langle \overline{I}\right\rangle _{\tau
_{c},\tau _{e}}^{2}\right)%
\end{array}%
\end{array}%
\end{equation*}%
where $\left\langle \overline{I}_{i}\overline{I}_{j}\right\rangle _{\tau
_{c},\tau _{e}}=\left\langle \overline{I}_{i}\right\rangle _{\tau _{c},\tau
_{e}}\left\langle \overline{I}_{j}\right\rangle _{\tau _{c},\tau
_{e}}=\left\langle \overline{I}\right\rangle _{\tau _{c},\tau _{e}}^{2}$.

So, this amount can be calculated:
\begin{equation*}
\begin{array}{lll}
\left\langle \overline{I}^{2}\right\rangle _{\tau _{c},\tau _{e}} & = &
\frac{\delta ^{2}}{2Q}\int\limits_{-Q}^{Q}\frac{\left( 1+e^{-q}\right) ^{2}}{%
4(1+\cosh q)^{2}}dq \\
&  &  \\
& = & \frac{\delta ^{2}}{(2Q)(e^{Q}+1)}\left( Q-e^{Q}+Qe^{Q}+1\right) \\
&  &  \\
& = & \delta ^{2}\frac{Q-\tanh (Q/2)}{2Q}.%
\end{array}%
\end{equation*}

Because $\sum\nolimits_{N_{tr}=0}^{\infty }N^{N_{tr}}N_{tr}\dfrac{e^{-_{N}}}{%
N_{tr}!}=N$ and $\sum\nolimits_{N_{tr}=0}^{\infty }N_{tr}(N_{tr}-1)\dfrac{%
e^{-_{N}}}{N_{tr}!}=N^{2}$, thus we find that

\begin{equation*}
\left\langle I^{2}\right\rangle =\frac{N\ \left\langle \delta
^{2}\right\rangle }{2Q}\left[ Q-\tanh (Q/2)\right] +\left\langle
I\right\rangle ^{2}
\end{equation*}%
what gives $var(I)=$ $\left\langle I^{2}\right\rangle -\left\langle
I\right\rangle ^{2}=\frac{N\ \left\langle \delta ^{2}\right\rangle }{2Q}%
\left[ Q-\tanh (Q/2)\right] .$ And thus the relative error is described by
\begin{equation}
\begin{array}{lll}
e(I) & = & \frac{\sqrt{var(I)}}{\left\langle I\right\rangle } \\
&  &  \\
& = & \left( 2\frac{\left\langle \delta ^{2}\right\rangle }{\left\langle
\delta \right\rangle ^{2}}\right) ^{1/2}N^{-1/2}\left[ \frac{Q-\tanh (Q/2)}{Q%
}\right] ^{1/2} \\
&  &  \\
& = & \left( 2\frac{\left\langle \delta ^{2}\right\rangle }{\left\langle
\delta \right\rangle ^{2}}\right) ^{1/2}\frac{(\ln 10\ N_{dec}W\ L)^{-1/2}}{%
(p_{\max }-p_{\min })^{1/2}}\left[ \frac{Q-\tanh (Q/2)}{Q}\right] ^{1/2} \\
&  &  \\
& = & K\left( \frac{Q-\tanh (Q/2)}{Q}\right) ^{1/2}%
\end{array}
\label{hiperbolic_equation}
\end{equation}%
where $K=\left( 2\frac{\left\langle \delta ^{2}\right\rangle }{\left\langle
\delta \right\rangle ^{2}}\right) ^{1/2}\frac{(\ln 10\ N_{dec}W\ L)^{-1/2}}{%
(p_{\max }-p_{\min })^{1/2}}$.

Thus by considering the variable $q=(E_{T}-\mu )/k_{B}T$ being uniformly
distributed, the relative error behaves as a simple logistic function of the
band gap amplitude ($Q$).

Aiming to propose a more realistic model then the supposition of the density
of states following an uniform distribution, we propose an alternative model
that considers the density of states as being quadratically distributed. The
theoretical quadratic function is a reasonable approximation to experimental
results \cite{WY1990} and is simpler than the polynomial fit to those
experimental data. For this, the theoretical $u-$shape modeled as a
quadratic function defined in the interval $[-Q,Q]$ is here proposed:

\begin{equation*}
\Pr (q)=aq^{2}+c.
\end{equation*}

Supposing normalization $\int_{-Q}^{Q}$ $\Pr (q)dq=1$, i.e., $%
2aQ^{3}/3+2cQ=1 $, we have
\begin{equation}
\Pr (q)=aq^{2}+\frac{(1-2aQ^{3}/3)}{2Q},  \label{theoretical_u_shape_curve}
\end{equation}%
and from the existence condition $c=(1-2aQ^{3}/3)>0$ and then $a<3/(2Q^{3})$

Following the same procedure used before (density of states
uniformly-distributed) now applied to the case which the density of states
follows a quadratic curve, we first calculate the first moment and then
compute the average of current considering the random telegraph noise caused
by all traps in the interface:
\begin{equation}
\begin{array}{lll}
\left\langle I\right\rangle & = & \left\langle \delta \right\rangle
N\int\limits_{-Q}^{Q}\frac{1+e^{-q}}{2(1+\cosh q)}\Pr (q)dq \\
&  &  \\
&  & \frac{\left\langle \delta \right\rangle N}{2}\int\limits_{-Q}^{Q}\frac{%
(1+e^{-q})\left[ aq^{2}+\frac{(1-2aQ^{3}/3)}{2Q}\right] }{(1+\cosh q)}dq%
\end{array}
\label{current}
\end{equation}

This result must be obtained by using numerical methods for each $a$. For $%
a\rightarrow 0$, we have the expected $\left\langle I\right\rangle
\rightarrow \frac{N\ \left\langle \delta \right\rangle }{2}$. We tested many
values $a<3/(2Q^{3})$ and no meaningful deviates of $\left\langle
I\right\rangle =\frac{N\ \left\langle \delta \right\rangle }{2}$ were
observed. Indeed just corrections of order $O(10^{-7})$ were verified, and
there are no evidences to believe there might be large deviations for the
current considering our theoretical u-shape curve. Our tests were performed
using $Q_{\max }=10$, and to perform it, just values $a<3/(2\cdot
10^{3})=0.0015$ were considered. We have employed the Simpsom rule (see for
example \cite{Press1986}) for the numeric integration of equation \ref%
{current}. The reader might think $a$ is very small and differences would be
observed at higher values of $a$. In order to clarify this issue, we
experimented $a=0.01$ up to$\ a=0.1874<3/(2\cdot 2^{3})=\allowbreak
0.187\,5\ $corresponding to $Q_{\max }=2$ and the same conclusion was
obtained: $\left\langle I\right\rangle $ is constant under $Q$ variations.
Thus, the relationship between $\left\langle I\right\rangle $ and $Q$ is
independent of $a$.

The most accurate model comes from the actual density of states, which can
be measured from experimental data as in \cite{WY1990} and fitted by a
polynomial. We have verified the differences in the MOSFET current
fluctuations between modeling the band-gap as a uniform distribution, a
theoretical u-shape modeled by the simple quadratic function described by
equation \ref{theoretical_u_shape_curve} and the actual experimental data
fitted by a $8^{th}$-degree polynomial.

The procedure of using the experimental density of states consists of
performing a scan of the figures of the experimental density of states found
in \cite{WY1990}, i.e., the density of traps having energy $E_{T}$,
corresponding to $q=(E_{T}-\mu )/k_{B}T$, here denoted by $f(q)$.

In that work 3 u-shape curves are studied, $f_{1}(q)\times q$, $%
f_{2}(q)\times q$ and, $f_{3}(q)\times q$, corresponding to 3 different
oxides (in the nomenclature of ref. \cite{WY1990}, reoxidized: oxide I,
nitrided: oxide II and, tceoxide: oxide III) defined in a symmetric band gap
interval $[-\alpha ,\alpha ]$. The first step was to map this band gap to an
interval $[-Q,Q]$, what was done with the simple transformation $q^{\prime
}=Q\left[ 1-\frac{(\alpha -q)}{\alpha }\right] $. For our scanning $n=100$
points, $\{\ \left( q_{i},\ f(q_{i})\right) \ \}_{i=1}^{n}$, were collected
from experimental figures and fitted by a 8-degree polynomial ($%
g_{i}(q^{\prime })=\sum\nolimits_{k=0}^{8}\beta _{k}q^{\prime \ k}$). The
degree of 8 was used because it presents a good visual fit for all u-shape
curves. A normalization is required, and thus $\Pr_{i}(q^{\prime
})=g_{i}(q^{\prime })/\int\nolimits_{-Q}^{Q}g_{i}(q^{\prime })dq^{\prime }=$
$\sum\nolimits_{k=1}^{8}\beta _{k}^{\prime }q^{\prime \ k}$, where $\beta
_{k}^{\prime }=\beta _{k}/\left( 2\sum\nolimits_{j=0}^{4}\beta
_{2j+1}(2j+1)^{-1}Q^{\ 2j+1}\right) $, for $q^{\prime }\in \lbrack -Q,Q]$.
And so a polynomial fit was performed for each value attributed to $Q$
changing from $Q_{\min }$ to $Q_{\max }$ and after we have the experimental
probability distribution describing the density of states. Next we perform
numerical integrations of the equations for the current and the relative
error obtained in this manuscript, which are performed using Simpsom method
with a very small step ($h\sim 10^{-3}$). It supplies high precision for the
integration and so for our estimates. These simple procedures are
implemented in Fortran.

%by{\LARGE \ space
%  reasons (LUCAS ISSO\ CABE\ AQUI - OU COLOCAMOS\ UM\ SIMPLES\
%  ALGORITMO DESCREVENDO\ ESTES PASSOS)}.
%For considering these functions as usual probability
%densities equation \ref{current} instead of using the theoretical u-shape
%given by equation \ref{theoretical_u_shape_curve}.

\begin{figure}[th]
\centering\includegraphics[width=0.9\columnwidth]{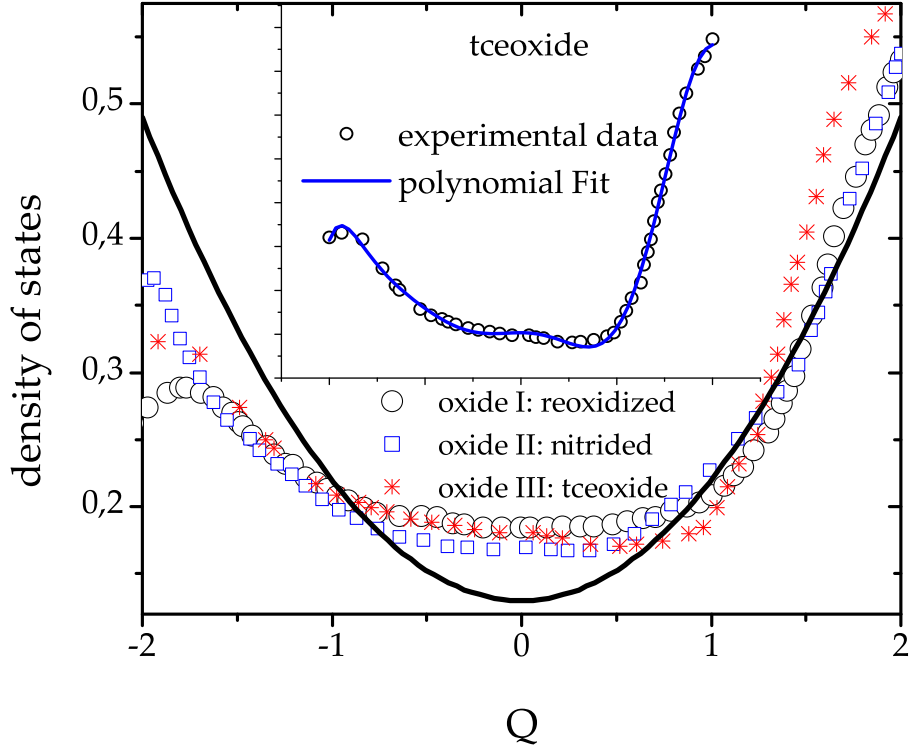}
\caption{Experimental density of states (U-shape curves) for 3 different
oxides. The data was extracted from ref. \protect\cite{WY1990}. The inside
plot corresponds to a example of the good polynomial fit by a nine degree
polynomial for the oxide III (tceoxide in the nomenclature of \protect\cite%
{WY1990}). Just for a comparisom the continuous curve represents a
theoretical curve for $a=0.09$ for the same $Q=2$.}
\label{plot_experimental_density_of_states}
\end{figure}

Figure \ref{plot_experimental_density_of_states} shows the experimental
u-shapes. The inside figure corresponds to an example of good polynomial fit
performed for the one of the oxides (tceoxide). So we perform calculations
to obtain $\left\langle I\right\rangle $.

\begin{figure}[th]
\centering\includegraphics[width=0.9\columnwidth]{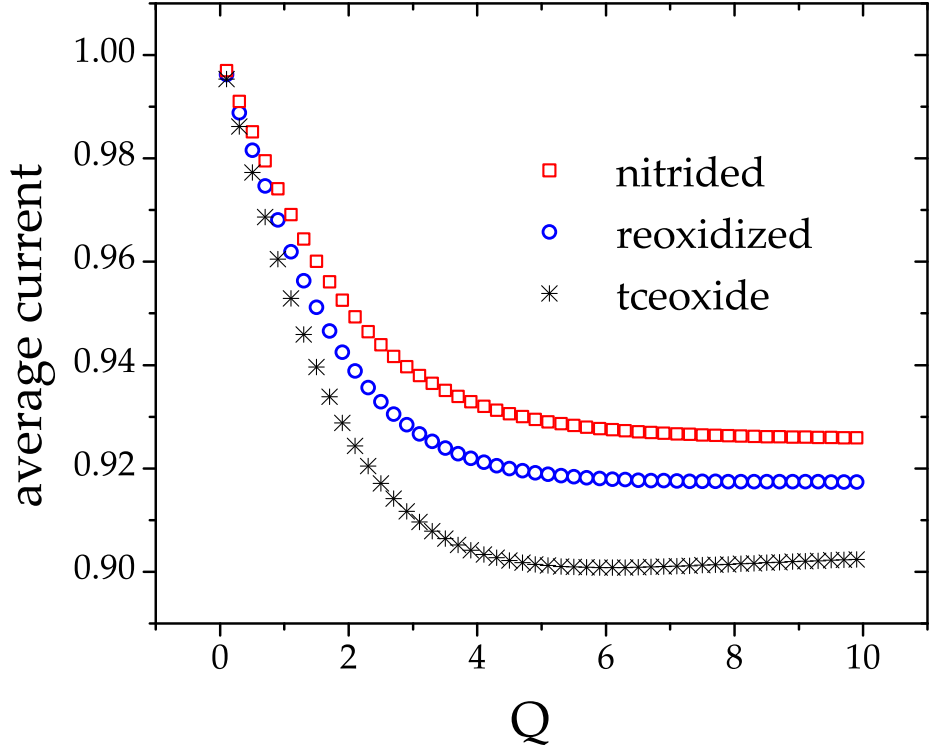}
\caption{Average current using the 3 experimental density of states shown in
figure \protect\ref{plot_experimental_density_of_states} as function of $Q$.
The results show a decay which is not observed when the density of states is
uniform or even when our theoretical u-shape density of states is
considered. }
\label{current_experimental}
\end{figure}

In figure \ref{current_experimental} we can observe that $\left\langle
I\right\rangle \ $ decays as function of $Q$ which is not observed when the
density of states is uniform or even when the curve is the theoretical
u-shape. This corroborates the idea that the asymmetry of the u-shape might
be important to observe the current differences that appear when $Q$
changes. However, it is important to notice that the average noise current
decays at most $10\%$ (for the tceoxide), for high values of Q. Since Q for S%
$_{i}$O$_{2}$ is approximately 2, in this case the decrease is nearly 5\%.
It is also important to analyze the relative error in the current by
computing its higher moments.

We then calculate the second moment for the theoretical u-shape density:%
\begin{equation*}
\left\langle \overline{I}^{2}\right\rangle _{\tau _{c},\tau
_{e}}=\left\langle \delta ^{2}\right\rangle \int\limits_{-Q}^{Q}\frac{%
(1+e^{-q})^{2}\left[ aq^{2}+\frac{(1-2aQ^{3}/3)}{2Q}\right] }{4(1+\cosh
q)^{2}}dq
\end{equation*}%
and the variance is
\begin{equation*}
var(I)=N\left\langle \delta ^{2}\right\rangle \int\limits_{-Q}^{Q}\frac{%
(1+e^{-q})^{2}\left[ aq^{2}+\frac{(1-2aQ^{3}/3)}{2Q}\right] }{4(1+\cosh
q)^{2}}dq\text{.}
\end{equation*}

\begin{figure}[th]
\centering\includegraphics[width=0.5\columnwidth]{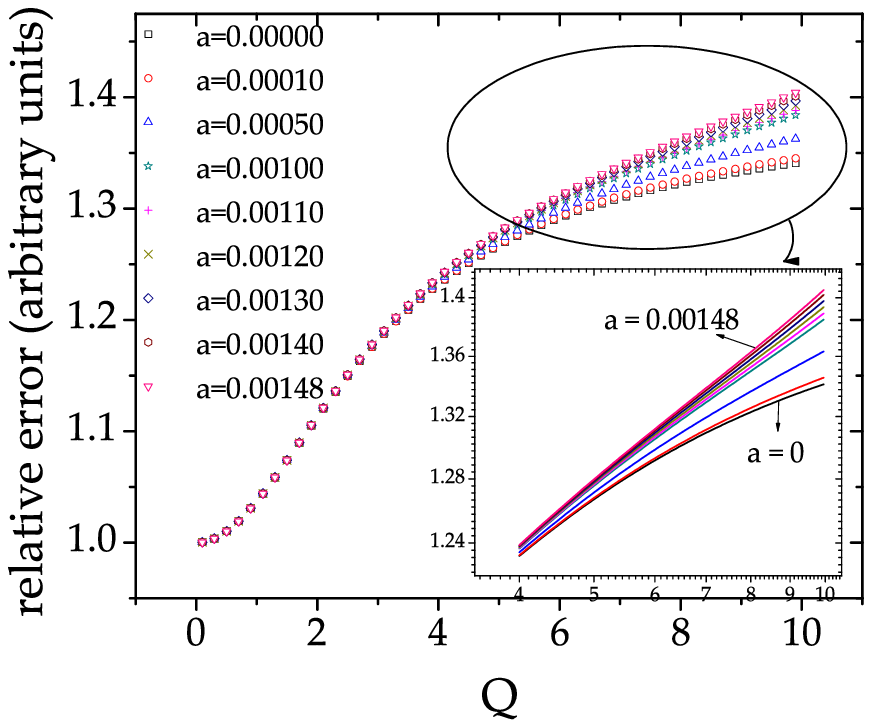}%
\includegraphics[width=0.5\columnwidth]{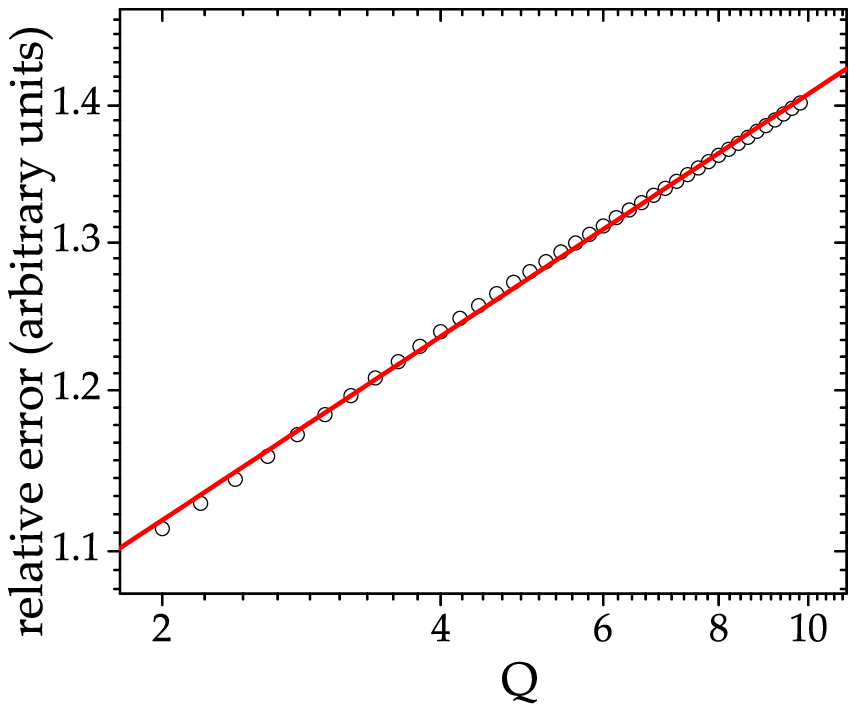}
\caption{\textbf{Left plot}: Relative error as function of $Q$ for a
u-shape(quadratic) density of states. The inside plot shows a log-log plot
for $Q>4$ showing that higher values of $a$ leads to a power law behavior of
the relative error as function of $Q$. \textbf{Right plot }A power law fit
of relative error as funcion of $Q$. It was used $a=0.00148$ and $Q>2$ for
the plots.}
\label{quadratic_density}
\end{figure}

We can calculate also in this case the relative error after a few algebraic
manipulation we have
\begin{equation*}
e(I)=K\frac{\ \left( \int\limits_{-Q}^{Q}\frac{(1+e^{-q})^{2}\left[ aq^{2}+%
\frac{(1-2aQ^{3}/3)}{2Q}\right] }{(1+\cosh q)^{2}}dq\right) ^{1/2}}{%
\int\limits_{-Q}^{Q}\frac{(1+e^{-q})\left[ aq^{2}+\frac{(1-2aQ^{3}/3)}{2Q}%
\right] }{(1+\cosh q)}dq}
\end{equation*}%
where $K=\left( 2\frac{\left\langle \delta ^{2}\right\rangle }{\left\langle
\delta \right\rangle ^{2}}\right) ^{1/2}\frac{(\ln 10\ N_{dec}W\ L)^{-1/2}}{%
(p_{\max }-p_{\min })^{1/2}}$. Integrating numerically the last equation we
have in the left plot of figure \ref{quadratic_density} the behavior of $e(I)
$ as function of $Q$, for different value of $a<0.0015$ showing that the
relative error is sensitive for different and small (possible) values of $a$
. For a suitable comparison, it is plotted in the same figure the case $a=0$%
, which corresponds to the uniform distribution of density of states. The
inside plot shows the same plot in a log-log scale. For higher values of $a$
a power law behavior can be adopted to describe the $e(I)\times $ $Q$. The
left plot of the same figure \ref{quadratic_density} describes that the
power law fit $e(I)=aQ^{\theta }$ considering $Q>2$ is suitable simple fit
for this interval. We find the best estimate $\theta =0.143(1)$ and $%
a=1.013(1)$.

These results show that the relative error in the current is not universal
since the relative error shows different behavior depending on the density
of states. Finally we have performed the same calculations for the relative
error considering the experimental u-shapes, as shown in figure \ref%
{experimental_relative_error}. We observe a difference in the dependence of $%
Q$ between the 3 different u-shapes. A similar behavior (logistic curve),
predicted for the uniform case (see equation \ref{hiperbolic_equation}), is
observed using the experimental u-shapes, however there is a sensitive
difference for higher values of $Q$ between the two cases. Our theoretical
approach for the u-shape supplies exactly an intermediate behavior between
the uniform and experimental u-shape results, showing it is suitable for
modeling RTS noise in MOSFETs because it is simpler then the higher-order
polynomial fit and more accurate than the uniform case!

\begin{figure}[th]
\centering\includegraphics[width=0.9\columnwidth]{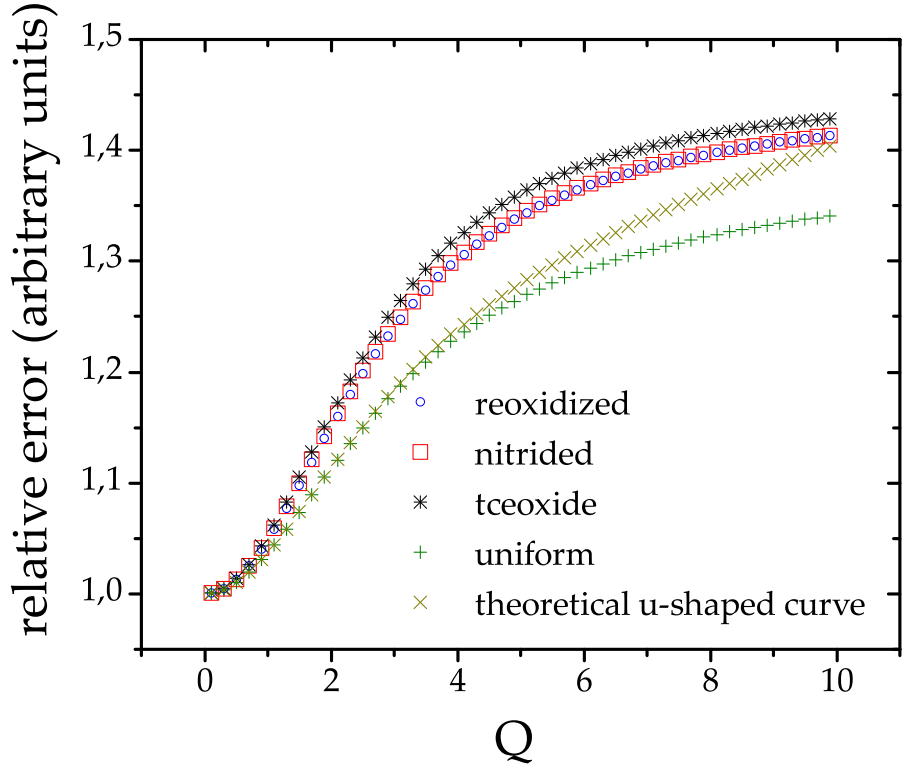}
\caption{Relative error as function of $Q$ for the experimental u-shapes.}
\label{experimental_relative_error}
\end{figure}

Summarizing, we developed an original methodology in time domain to model
the noise behavior due to traps at the S$_{i}$-S$_{i}$O$_{2}$ interface. Our
results show that the relative error depends on the amplitude ($Q$) of the
density of states at the interface.

Our model for the average current and its relative error depends upon the
model of the density of states adopted. When considering the density of
states as having uniform or quadratic density of trap energies in the band
gap we conclude that the relative error behaves as a logistic function in
relation to Q. For the density of states modeled as an uniform distribution,
a very simple formulation can still capture a logistic behavior similar to
the experimental data.

The experimental u-shapes show a decay in the average current not observed
when the theoretical u-shape is considered. Such results represent an
important advance in modelling RTS at device and circuit levels, and
experiments can be performed to check these behaviors. We believe that
device designers can use the models proposed in this work for modeling the
relative error as a function of the band-gap in upcoming technologies.

%\biliography{bsim}

\section*{Acknowledgments}

We would like to thank the anonymous referees for the helpful
suggestions. R. da Silva and G. I. Wirth are financially supported
by the following projects of CNPq: 490440/2007, 480258/2008-2,
311343/2006-6, and 577473/2008-5.

\end{document}